\documentclass[amsmath,superscriptaddress,showpacs,aps,prl,twocolumn]{revtex4-1}
\usepackage[colorlinks,linkcolor=blue,anchorcolor=blue,citecolor=blue,urlcolor=blue]{hyperref}
\usepackage{amsmath}
\usepackage{amssymb}
\usepackage{graphicx}
\usepackage{color}
\usepackage{bm}

\begin{document}
\title{Confinement and oscillating deconfinement crossover of two-magnon excitations in quantum spin chains quantified by spin entanglement entropy
}

\author{Zhao-Yang Dong}
\affiliation{Department of Applied Physics, Nanjing University of Science and Technology, Nanjing 210094, China.}
\author{Jian-Xin Li}
\email[]{jxli@nju.edu.cn}
\affiliation{National Laboratory of Solid State Microstructures and Department of Physics, Nanjing University, Nanjing 210093, China}
\affiliation{Collaborative Innovation Center of Advanced Microstructures, Nanjing University, Nanjing 210093, China}
\date{\today}

\begin{abstract}

\par We introduce the spin entanglement entropy (EE) to characterize spin excitations. The scalings of EEs are elaborated as $\ln N$, $\ln N+D_k$ and $2\ln N+D$ for the single-magnon states, two-magnon bound states and two-magnon continuum, respectively, based on the Bethe ansatz solutions of the ferromagnetic spin chain with $N$ sites. The $\ln N$ divergence for the bound states reveals the two magnons emerge as a new quasiparticle. More importantly, the nonzero intercepts ($D_k,D$) embody the many-body effects of two-magnon states. In particular, an exact relation between the intercept of EEs and an observable quantity, the two-magnon distance, is established for the bound states. In such a case, we quantify the two-magnon confinement in a spin system by the increasing entanglement with the distance as a particle physics analogue. Moreover, the EEs can also be used to study the evolution of the excitations. When the bound states are immersed in the continuum in the alternating chain, they undergo an oscillating confinement-deconfinement crossover, shown by the oscillations of the EEs and two-magnon distance with the chain length.



\end{abstract}

\maketitle

\par The confinement describes the phenomenon that the constituent particles are bound together by an interaction which strength increases with particle separation, which most prominently found in quantum chromodynamics between quarks \cite{PhysRevD.10.2445}. As an analogue, it has also been suggested to reflects the intrinsic many-body features of quantum spin systems.
Established cases are the confinement of spinons (domain walls) illustrated by a potential resulted from frustrated rungs between the spinons in the spin ladders \cite{Lake2010,Greiter2010} or a magnetic field \cite{Kormos2017,Tan2021}, and confinement of chargons on Ising chains \cite{PhysRevLett.124.120503,PhysRevLett.125.256401,PhysRevLett.127.167203}.
\par In a spin $s=1/2$ ferromagnetic Heisenberg chain, the celebrated Bethe ansatz solution showed there are two-magnon bound states, in which magnons are bound as a single emergent entity \cite{Bethe1931}. The two-magnon bound state has also been acknowledged in various systems with arbitrary spin, such as 2D and 3D ferromagnetic and antiferromagnetic system \cite{PhysRev.102.1217,LOVESEY196984,PhysRevLett.11.336,PhysRev.132.85, Haldane_1982,Southern_1994,PhysRevLett.19.30,10.1143/PTPS.46.61,10.1143/PTP.46.401,doi:10.1139/p72-234,doi:10.1143/JPSJ.30.358,doi:10.1063/1.1664749,PhysRevB.7.2207, doi:10.1143/JPSJ.34.1486,PhysRevB.9.4939,doi:10.1143/JPSJ.31.394,Fogedby_1980,PhysRevB.24.5327,PhysRevB.76.060407,PhysRevB.96.195134,Qin_2018,PhysRevA.97.043415},
and detected experimentally \cite{PhysRev.187.595,PhysRevLett.87.127002,PhysRevLett.108.077206,Fukuhara2013,PhysRevB.101.180401,PhysRevLett.118.177202,PhysRevLett.125.087202}.
In the exact Bethe ansatz solution, two-particle excitations are evidenced as bound states if the wavefunction is localized, or as individual excitations consisting of two almost free magnons if spatially extended.
However, how to characterize the confinement of magnons in the bound states is an open issue.
On the other hand, considering that quasiparticles inevitably become unstable when encountering the continuum of many-particle excited states, a confinement-deconfinement transition is expected when there is an interplay between two-magnon bound states and continuum.
So far, the fate of the bound states with this interplay is usually studied by the evolution of the spectral line shape from sharp Lorentzian peaks (bound states) to a broad continuum. The bound states are often considered to be destabilized when swallowed by the continuum, or may be stabilized by strong interactions pushing them out of the continuum \cite{Verresen2019}. It still remains elusive to quantify the evolution of the bound states and uncover the nature of the transition.



\par Confinement leads to enhanced quantum coherence \cite{Konik2021}, referring to quantum entanglement which is increasingly considered as a vital resource to our understanding of the many-body systems~\cite{RevModPhys.80.517,Vedral2008,PhysRevLett.90.227902,RevModPhys.89.041004,Broholm2020,Jiang2012,PhysRevLett.105.116805}.
One way of quantifying the entanglement between two regions of a system is through the entanglement entropy. Real space bipartitions have been usually used and extensively studied~\cite{RevModPhys.82.277,LAFLORENCIE20161}. In this letter, we introduce the spin entanglement entropy (EE) \cite{Dong2020,PhysRevB.104.L180406} to characterize spin excitations and the confinement of the bound states. The spin EE is based on the bipartition of a many-body quantum system into two different spin regions: region $A$ describes the spin-up states and region $B$ the spin-down states. In such a way, the wavefunction $|\psi\rangle$ of the system with $N$ spins is written as $|\psi\rangle=\sum_{\{n\}}a(\{n\})|\{n\}\rangle_\uparrow\otimes|\{N\}-\{n\}\rangle_\downarrow$, where $|\{n\}\rangle_\uparrow$, in the region $A$, indicates the configuration of $n$ spin-up sites, and $|\{N\}-\{n\}\rangle_\downarrow$, in the region $B$, the configuration of $N-n$ spin-down sites.
Therefore, the spin EE can be defined between the spin-up and down spaces due to the Schmidt decomposition of the wavefunctions with respect to this bipartition,
\begin{equation}\label{EE}
  S=\sum_{\{n\}}-a(\{n\})^2 {\rm ln}(a(\{n\})^2),
\end{equation}
and its difference from the ground state can be used to analysis spin excitations. For instance, 1-dimensional system can be viewed as a two leg ladder picturely, with the upper leg representing spin up, and the lower leg spin down. Entanglement entropy is then defined by partitioning the system into these two legs. As one of the intrinsic attributes, the scalings of spin EEs could distinguish different spin entanglements of different spin excitations. For example, a spin-flipping excitation can be decomposed into two spin-1/2 spinons and then be studied by our spin EE. If the two spinons are confined to form a particle-hole pair entangled at every site, the spin-flipping excitation is identified as a magnon whose spin EE will be logarithmically divergent with the system size $N$ in the thermodynamic limit \cite{Dong2020,PhysRevB.104.L180406}. Otherwise, if the spin-1 excitation are fractionalized into individual spin-1/2 spinons as realized in a quantum spin liquid, the spin EE converges to a constant. In the same way, to character the $n$-magnon states we can consider $n$ spin-flipping excitations and analysis the corresponding scaling behavior of the spin EE.

\par Firstly, we will elaborate the EEs based on the Bethe ansatz solutions of spin excitations in a ferromagnetic Heisenberg chain with $N$ sites. The scalings of EEs of single-magnon states, two-magnon bound states and continuum diverge as $\ln N$, $\ln N+D_k$ and $2 \ln N+D$, respectively.
In contrast to the single-magnon state, the nonzero intercepts for the bound states and continuum embody the many-body effects between magnons.
In particular, an exact relation between the intercept of the EEs and two-magnon distance for the bound state is established.
According to the relation, the monotonously increasing entanglement with the distance between two magnons illustrates the two-magnon confinement in the bound states quantitatively. While, the negative EE intercept for the two-magnon continuum reflects the indistinguishability and coherence of individual magnons.
To study the evolution of the two-magnon bound states in the continuum, we turn to the alternating $J-J'$ ferromagnetic Heisenberg chain. From the spectral perspective, only a gradual suppression of the bound states is observed when they encounter the continuum.
In sharp contrast, our EE analysis unveils an oscillation of both the EEs and the distance between magnons with the system size. This unusual oscillation suggests that the bound states are neither deconfined to be individual magnons nor preserved from the scattering. Instead, they experience an oscillating confinement-deconfinement crossover.




\par The Hamiltonian of the spin-$1/2$ Heisenberg model on a $N$-site chain is written as $H=-J\sum_i \mathbf{S}_i\cdot\mathbf{S}_{i+1}$.  The ground state is ferromagnetic ordered, $|F\rangle=|\uparrow\ldots\uparrow\rangle$ (all spins up) with energy $E_0=-JN/4$. Obviously, with respect to the bipartition in spin space, the EE of the ground state is zero. The spin-$1$ elementary excitations, known as magnons, can be obtained by diagonalizing $H$ in the space with one spin flips $|r\rangle=S_r^-|F\rangle$ ($r$ labels the lattice site). The eigenstates are given by $|o_k\rangle={1 \over \sqrt{N}}\sum_r e^{ikr} |r\rangle $,
where $k=2\pi\lambda/N$ is the wave vector, and the eigenvalue is $E_k=J(1- \cos k)$. The EE of magnons is $\ln N$, and its perfectly logarithmic divergence without intercept aptly illustrates the single entity property of a magnon.

\par If there are two free magnons, the EE is expected to be $2\ln N$, which is a simple sum of that of each magnon. To check this, we analyse the two-magnon excitations in the space with two spins flip $|r_1,r_2\rangle=S^-_{r_1}S^-_{r_2}|F\rangle$. The eigenstates are given by the Bethe ansatz $|t_k\rangle=\sum_{r_1<r_2}\psi(r_1,r_2)|r_1,r_2\rangle$ with the coefficients \cite{Bethe1931,Karabach1997}
\begin{equation}\label{two}
  \psi(r_1,r_2)=e^{ i(k_1r_1+k_2r_2+{\theta \over 2})}+e^{ i(k_1r_2+k_2r_1-{\theta \over 2})},
\end{equation}
where $k_1,k_2,\theta$ satisfy $2\cot{(\theta/2)}=\cot{(k_1/2)}-\cot {(k_2/2)}$.
The periodic boundary conditions require $Nk_1=2\pi\lambda_1+\theta$ and $Nk_2=2\pi\lambda_2-\theta$, where $\lambda_i$ are known as Bethe quantum numbers.
Due to the translational symmetry of the chain, the wave vectors $k=k_1+k_2=2\pi(\lambda_1+\lambda_2)/N$ are still good quantum numbers. The magnon interaction is reflected in the phase shift $\theta$ and the deviation of the momenta $k_1, k_2$ from the values of the one-magnon wave vectors. Two classes of two-magnon excitations are classified according to the Bethe quantum numbers, i.e., two-magnon bound states and continuum. 

\begin{figure}
\centering
\includegraphics[width=0.45\textwidth]{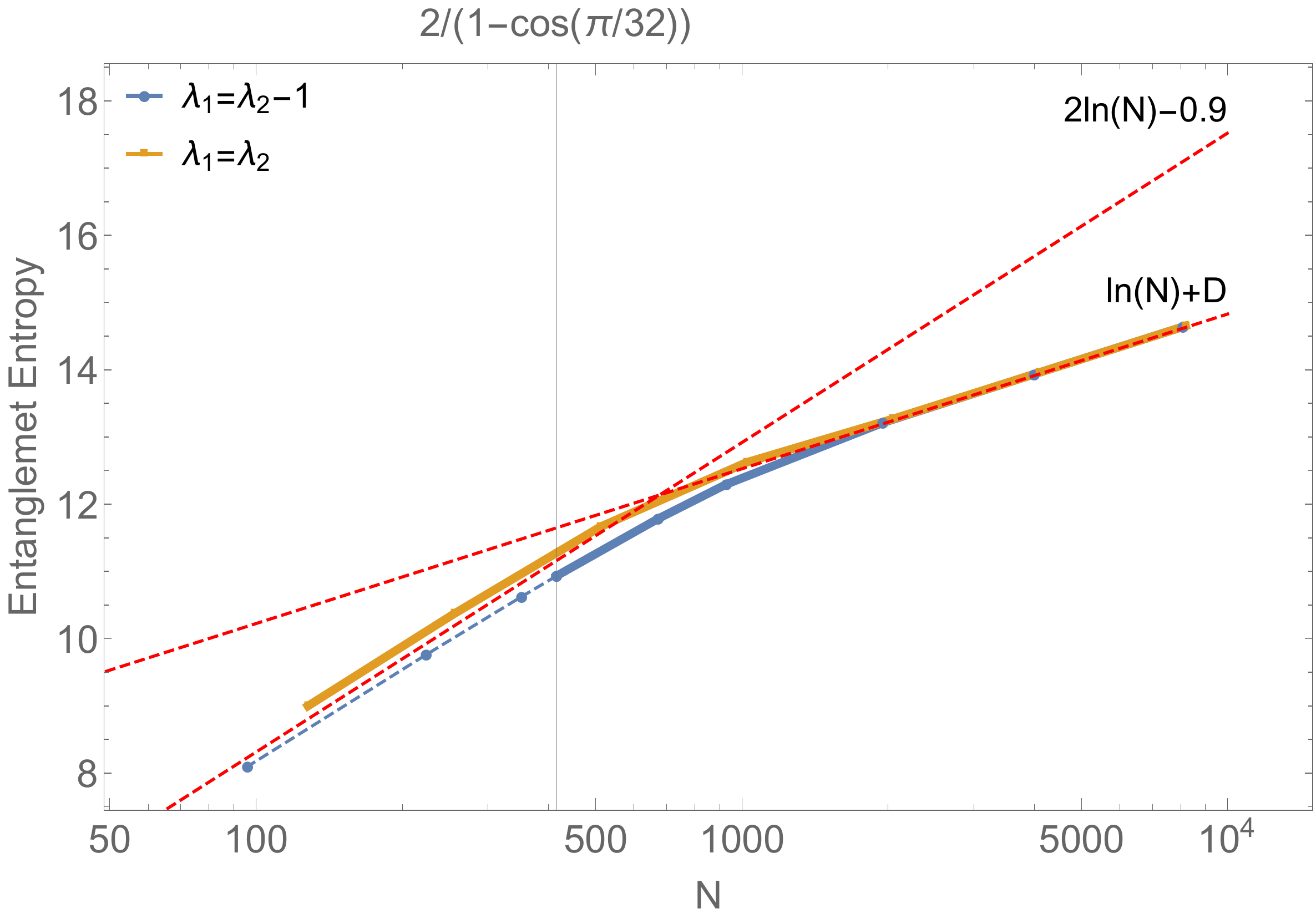}
\caption{(color online). 
Scaling of EEs of the two-magnon bound state at $k=\pi/16$, plotted in log-linear scale, where two kinds of solutions are denoted as different colors. The blue dashed line for $\lambda_1=\lambda_2-1$ indicates the real $k_1, k_2, \theta$ solutions. The EEs of both kinds of solutions for the bound state approach to $\ln N+D_k$ in the thermodynamic limit, but approach to $2\ln N-0.9$ in the finite-size chain when $N<2/(1-\cos(\pi/32))$. }
\label{analysis}
\end{figure}

\par {\it Bound states.--} When $\lambda_1-\lambda_2\leq1$, $k_1, k_2, \theta$ are complex, $k_1=k/2+iv,k_2=k/2-iv,\theta=\pi(\lambda_1-\lambda_2)+iNv$. The complex solutions are supposed to result in localized wavefunctions, which represents two magnons will form the two-magnon bound states. Their spin EEs are,
\begin{equation}\label{EE1}
  S_k=\ln \mathfrak{N}-\sum_{r_1<r_2}{p_{r_1,r_2} \over \mathfrak{N}},
\end{equation}
and $\mathfrak{N}$ is the normalization constant of Eq.~(\ref{two}),
\begin{eqnarray}
  \mathfrak{N} &=& \sum_{r_1<r_2} p_{r_1,r_2}= \sum_{r_1<r_2}|\psi(r_1,r_2)|^2, \nonumber\\
  |\psi(r_1,r_2)|^2  &=& \{\begin{array}{l}
            2(\cosh [(N-2 \Delta r)v]+1),~\lambda_1=\lambda_2\\
            2(\cosh [(N-2 \Delta r)v]-1),~\lambda_1=\lambda_2-1 \nonumber
          \end{array},
\end{eqnarray}
where $\Delta r=r_2-r_1$. For a given wave vector $k$, there are two kinds of solutions for different $N$. Fortunately, they are unified as a single branch with dispersion $E_k=J(1-\cos k)/2$ in the thermodynamic limit, and $\lim_{N\rightarrow\infty}\mathfrak{N}=N$, $\lim_{N\rightarrow\infty} p=q^{2\Delta r-2}-q^{2\Delta r}$, where $q=e^{-v}=\cos (k/2)$. Therefore, in the thermodynamic limit, the EEs of bound states read,
\begin{eqnarray}
    S_k&=&\ln N+D_k, \label{boundEE}\\
  D_k&=&-\ln (1-q^2)-{q^2 \over 1-q^2} \ln q^2. \label{Dk}
\end{eqnarray}
The $\ln N$ divergence of the EEs indicates that a single new quasiparticle emerges in place of two magnons.
Curiously, an unusual nonzero intercept $D_k$ suggests that the bound state behaves more than a single entity of the quasiparticle.
Reminding the distance between the two magnons in the bound state is $d_k=\lim_{N\rightarrow\infty}\langle t_k|\Delta r|t_k\rangle=1/(1-q^2)$, we can obtain an exact relation between $D_k$ and $d_k$ by substituting $d_k$ into Eq.~(\ref{Dk}),
\begin{equation}\label{Dkr}
  D_k=d_k\ln d_k-(d_k-1)\ln (d_k-1).
\end{equation}
A remarkable property that Eq.~(\ref{Dkr}) implies is the entanglement between two magnons is enhanced monotonously with their distance. It is known that the intercepts, which act as corrections to the $\ln N$ divergent EE, would reflect the effects of the interactions between magnons. By analogy with the behavior of interactions in the quark confinement, we may suggest that the nonzero intercept of EEs might be contributed from gluon-like excitations, which are active in spin space to confine the two magnons. When $k=\pi$, where the bound state has the highest energy, one gets $d_k=1$ from $d_k=1/(1-q^2)$, suggesting there is no space between the two magnons for active gluon-like excitations. In this case, that the gluon-like excitations are also confined in the entity leads to $D_k=0$, and this bound state will behave as a single emergent quasiparticle, whose EE exhibits a $\ln N$ scaling.
As $k$ decreases, $d_k$ and $D_k$ increases along with the decline of the energy consequently. In other words, the two magnons in the bound state would entangle more weakly when they were bound more tightly at sufficiently high energies, 
which is similar to the concept of asymptotic freedom in particle physics. Therefore, though in the one-dimensional chain with only nearest-neighbour couplings, the increasing intercepts with the distance suggest that the possible gluon-like excitations could confine the two magnons far away from each other.
Eventually, $d_k$ and $D_k$ are divergent at $k=0$, implying deconfinement of the bound states. These correspond to two solutions in which $k_1, k_2$ are reals and $\theta=0$, i.e., $k_1=k_2=0$ and $k_1=0, k_2=2\pi/N$. Their wavefunctions are extended and corresponding EEs are $2\ln N-\ln2$ and $2\ln N-1$ in the thermodynamic limit, respectively. The $2\ln N$ divergence illustrates that the two-magnon states are two individual magons rather than confined bound states, and the nonzero intercepts represent the indistinguishability and coherence of magnons.

\par As we known, for a finite-size system with a given $k$, there are distinctions between the two kinds of solutions depending on the system size $N$. When $N<2/(1-|q|)$, that the solutions for $\lambda_1=\lambda_2-1$ are real but for $\lambda_1=\lambda_2$ are complex suggests the former describes two individual magnons, while the latter the bound state. However, their EEs behave in the same way. 
As an example, the scaling of EEs for the bound state at $k=\pi/16$ are shown in Fig.~\ref{analysis}. 
When $N<2/(1-\cos(\pi/32))$, the EEs of both solutions follow $2\ln N$ scaling that the bound state is deconfined to be two individual magnons in a finite-size system no matter whether the solution is real or not. While $N>2/(1-\cos(\pi/32))$, the EEs both approach to $\ln N+D_k$ in the thermodynamic limit as expected.





\par
\begin{figure}
\centering
\includegraphics[width=0.5\textwidth]{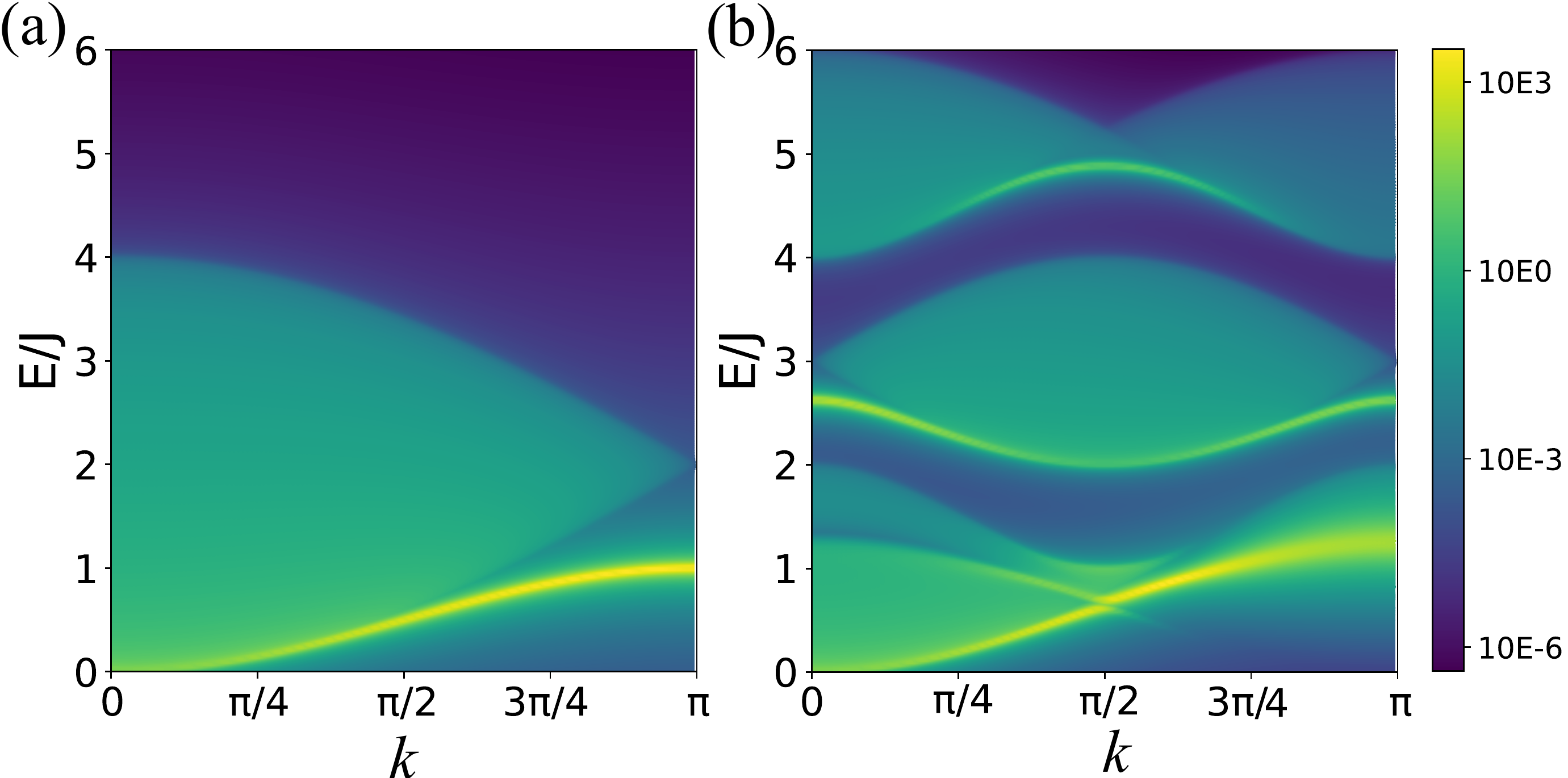}
\caption{(color online). Magnon pair correlation functions for the alternating Heisenberg chain, when (a) $J'=J$ and (b) $J'=2J$. }
\label{mpcspectra}
\end{figure}

 {\it Two-magnon continuum.--} The other class of solutions belongs to $\lambda_2-\lambda_1>1$, in which $k_1, k_2, \theta$ are all real. In this case, the wavefunctions are spatially extended, giving rise to a pair of magnons with energy $\varepsilon_{k_i}=J(1-\cos k_i)$. Therefore, the excitations result in a continuum $E_{k}=\varepsilon_{k_1}+\varepsilon_{k_2}$ beyond the bound states in the spectra as shown in Fig.~\ref{mpcspectra} (a). The corresponding EEs are also in the form of $\ln \mathfrak{N}+D$. However, $\lim_{N\rightarrow\infty}\mathfrak{N}=N^2$ gives a $2 \ln N$ divergent EE, the standard EE for two free magnons, which implies the states in the continuum is deconfined.
Moreover, the nonzero intercepts $D$, which almost converge to $-1$ (rarely distribute between $-2 \ln 2$ and $-\ln 2$) \cite{SM}, reflects the indistinguishability and coherence of individual magnons. That $-\ln 2$ originates from the denominator in the dimension $N(N-1)/2$ of Hilbert space of identical particles, and the part more than $-\ln 2$ results from the interference of wavefunctions illustrate the magnons as quantum particles. For examples, one of the strongest interfering solutions, $k_1=0,k_2=\pi,\theta=0$, results in $D=-2 \ln 2$. 
Besides, in the thermodynamical limit with periodic boundary conditions, the average distance between two magnons in the continuum is always $N/4$, which is irrelevant to the intercepts \cite{SM}.



\begin{figure*}
\centering
\includegraphics[width=0.99\textwidth]{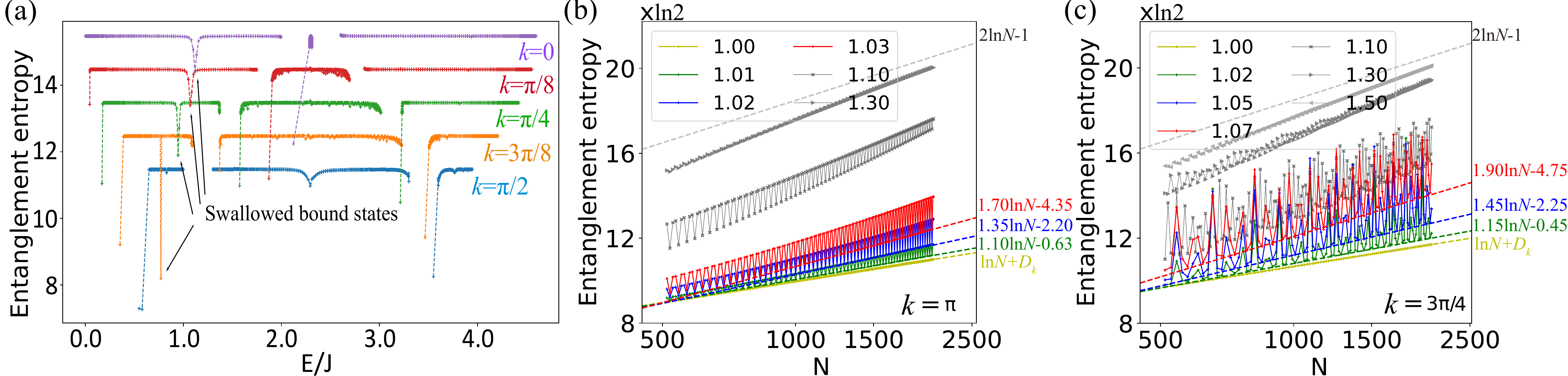}
\caption{(color online). (a) EE spectra of two-magnon excitations of $N=256$ chain with $J'/J=1.3$. The spectra for different $k$ are shifted by $1$ in the vertical direction for clarity. Scalings of EEs of the swallowed bound states with different $J'/J$ at (b) $k=\pi$ and (c) $k=3\pi/4$. Different $J'/J$ are denoted by different colors as indicated in legends. The grey dashed line denotes the reference line for the EEs of two-magnon continuum $2\ln N-1$.}
\label{eespectra}
\end{figure*}

\par {\it Evolution of bound states.--}
To introduce couplings between bound states and continuum, we turn to the alternating $J-J'$ Heisenberg chain,
\begin{equation}\label{H}
  H=-J\sum_{i\in odd} \mathbf{S}_i\cdot\mathbf{S}_{i+1}-J'\sum_{i\in even} \mathbf{S}_i\cdot\mathbf{S}_{i+1}.
\end{equation}
Since the Brillouin zone is folded as the translational symmetry is doubled, the folded continuum will overlap the two-magnon bound states to result in the couplings.
Let us first calculate the magnon pair correlation functions to have a look at the excitation spectra,
\begin{equation}\label{mpc}
  P(\omega,k)=-{1 \over N^2\pi}\sum_r{\rm Im}[\sum_{t_k} {|\langle F| S_r^+S_{r+1}^+|t_k\rangle|^2\over \omega-E_{k}+i0^+}]e^{i kr},
\end{equation}
where $|t_k\rangle$ and corresponding $E_{k}$ are solved by exact diagonalization. The results for $J'=J$ and $J'=2J$ are shown in Fig.~\ref{mpcspectra} (a) and (b), respectively. For $J'=J$, which has been discussed analytically above, a clear bound state dispersion shows up with the highest spectral weight below the two-magnon continuum. 
For $J'=2J$, the magnons evolve into optical and acoustic branches, so the two-magnon excitations appear as acoustic-acoustic, acoustic-optical and optical-optical excitations, exhibiting three parts separated by gaps as shown in Fig.~\ref{mpcspectra} (b).
Concurrently, the high-energy bound states encounter the folded acoustic-acoustic continuum, and an evolution arises with the ratio $J'/J$.
From the spectra as illustrated in Fig.~\ref{mpcspectra}, we can only observe the gradual suppression of the bound state peaks, but can distinguish neither the deconfinement of the bound states, nor the feature of the transition. Therefore, we resort to the EEs analysis to answer these questions.

\par The EE spectra of two-magnon excitations are shown in Fig.~\ref{eespectra} (a). It is noted that the EE spectra are symmetric about $k=\pi/2$, while the excitation spectra are not due to the structure factors [Fig.~\ref{mpcspectra} (b)].
The spectra at a given $k$ exhibit three segments, in agreement with the three continua composed of different branches of magnons. Because the EEs of continua are around $2\ln N-1$, it results in a uniform flat spectra.
On the other hand, the EEs of bound states $\ln N+D_k$ are much smaller than $2\ln N+D$ of continua for a fixed system size, so the well-defined bound states can be unambiguously found with relatively small EEs at the bottom of each continuum.
This quantitative criterion can also be used to identify or locate the bound states immersed in the continuum, which are hardly distinguished from the continuum based on the excitation spectra shown in Fig.~\ref{mpcspectra}(b). As shown from the EE spectra in Fig.~\ref{eespectra} (a), one can clearly find ravines in the acoustic-acoustic continuum spectra to identify the swallowed bound states quantitatively.



\par Now, we turn to the evolution of the swallowed bound states immersed in the continuum with the ratio $J'/J$.
In Fig.~\ref{eespectra} (b) and (c), we show the scaling of EEs for these states at $k=\pi$ and $k=3\pi/4$. Interestingly, when $J'$ begins to deviates from $J$, rather than approaching to a definite logarithmic scaling, the EEs oscillate as a function of the system size.
It is noted that the different size dependence of different bound states cause a $k$-dependent period of the oscillations.
For examples, when the deviation is small, $T_\pi=6$ and $T_{3\pi/4}=22.72$ repectively \cite{SM}.
Moreover, a similar oscillation of the distance \cite{SM} between two magnons implies they are obviously not two individual magnons, whose distance is irrelevant to the EE.
This unusual behaviour indicates that the bound states can not be considered conventionally to be preserved as quasiparticles with damping or deconfined into two individual magnons after encountering the continuum. Nevertheless, to examine the evolutions of the bound states, the lower boundary of the oscillations for $J'/J<1.10$ can be fitted by a logarithmic dependence, $\beta\ln N+C$, as shown by the colored dashed lines in Fig.~\ref{eespectra} (b) and (c). In the fittings, $\beta$ increases gradually with $J'/J$ from $1$, which is the well-defined bound states correspond to. When $J'$ increases further, such as $J'/J\geq1.10$, the oscillations turn to fade away slowly. Eventually, they exhibit a tendency to approach to $2\ln N-1$, the normal scaling behaviour of two individual magnons.
In a word, the EEs and distance between two magnons are oscillating in this process until a relative large $J'$. That is, the deconfinement transition of the two-magnon bound states in the continuum is like a melting of glass that happens gradually and has no critical point. Therefore, we would ascribe it as a confinement-deconfinement crossover rather than a phase transition.
Additionally, considering that the EEs of bound states are related to the distance $d_k$ as mentioned before, this oscillation is expected to be probed by measuring the distance between two magnons.


\par In this letter, we introduce the spin entanglement entropy to characterize the two-magnon excitations, confinement, and confinement-deconfinement crossover in the ferromagnetic spin chains.
An exact relation is unveiled between the intercept of EEs and the distance between two magnons to quantify the confinement of bound states, and build a bridge between the entanglement and an observable quantity, which may lay a foundation for studying the quantum entanglement by observing the dynamics of the bound states microscopically \cite{PhysRevLett.108.077206,Fukuhara2013}. When the bound states encounter the continuum in the alternating chain, we reveal a gradual oscillating confinement-deconfinement crossover instead of a phase transition by showing an oscillating behavior of the EEs and distance. Our study provides a systematic quantification of the magnetic excitations and a direct visualization of the two-magnon confinement and deconfinement in quantum ferromagnetic spin chains, which may help us understand the quantum coherence based on confinement \cite{Konik2021}. We also would like to mention that the unusual bipartition and the corresponding EE proposed here could be generalized to study the excitations with different degrees of freedom, such as the spin and charge excitations in the Luttinger liquid \cite{doi:10.1063/1.1704046,PhysRevLett.45.1358} and Luther-Emery liquid \cite{PhysRevLett.33.589,PhysRevLett.34.1247,PhysRevB.71.045113}.

\begin{acknowledgments}
\par We are grateful to Alexander Seidel, Craig Roberts and Z.F. Cui for helpful discussions. This work was supported by National Key Projects for Research and Development of China (Grant No. 2021YFA1400400), the National Natural Science Foundation of China (Grants No. 11904170, 92165205), the Natural Science Foundation of Jiangsu Province, China (Grant No. BK20190436), and the Doctoral Program of Innovation and Entrepreneurship in Jiangsu Province.
\end{acknowledgments}

\bibliography{reference}
\end{document}